\documentclass[prl,twocolumn,preprintnumbers,superscriptaddress,amsmath,amssymb]{revtex4}

\usepackage{graphicx}

\usepackage{dcolumn}

\begin{document}

\title{Strong reduction of exciton-phonon coupling in high crystalline quality single-wall carbon nanotubes: a new insight into broadening mechanisms and exciton localization.}

\author{V. Ardizzone}
\affiliation{Laboratoire Aim\'e Cotton, ENS Cachan, CNRS, Universit\'e Paris Sud, bat 505 campus d'Orsay, 91405 Orsay cedex France}
\author{Y. Chassagneux}
\author{F. Vialla}
\affiliation{Laboratoire Pierre Aigrain, Ecole Normale Sup\'erieure,UPMC, Universit\'e Paris Diderot, CNRS, 24 rue Lhomond 75005 Paris, France}
\author{G. Delport}
\author{C. Delcamp}
\affiliation{Laboratoire Aim\'e Cotton, ENS Cachan, CNRS, Universit\'e Paris Sud, bat 505 campus d'Orsay, 91405 Orsay cedex France}
\author{N. Belabas}
\affiliation{Laboratoire de Photonique et Nanostructure, CNRS, 91460 Marcoussis,France}
\author{E. Deleporte}
\affiliation{Laboratoire Aim\'e Cotton, ENS Cachan, CNRS, Universit\'e Paris Sud, bat 505 campus d'Orsay, 91405 Orsay cedex France}
\author{Ph. Roussignol}
\affiliation{Laboratoire Pierre Aigrain, Ecole Normale Sup\'erieure,UPMC, Universit\'e Paris Diderot, CNRS, 24 rue Lhomond 75005 Paris, France}
\author{I. Robert-Philip}
\affiliation{Laboratoire de Photonique et Nanostructure, CNRS, 91460 Marcoussis,France}
\author{C. Voisin}
\affiliation{Laboratoire Pierre Aigrain, Ecole Normale Sup\'erieure,UPMC, Universit\'e Paris Diderot, CNRS, 24 rue Lhomond 75005 Paris, France}
\author{J.S. Lauret}
\affiliation{Laboratoire Aim\'e Cotton, ENS Cachan, CNRS, Universit\'e Paris Sud, bat 505 campus d'Orsay, 91405 Orsay cedex France}
\email{jean-sebastien.lauret@lac.u-psud.fr}

\begin{abstract}
Carbon nanotubes are quantum sources whose emission can be tuned at telecommunication wavelengths by choosing the diameter appropriately. Most applications require the smallest possible linewidth. Therefore, the study of the underlying dephasing mechanisms is of utmost interest. Here, we report on the low-temperature photoluminescence of high crystalline quality individual single-wall carbon nanotubes synthesized by laser ablation (L-SWNTs) and emitting at telecommunication wavelengths. A thorough statistical analysis of their emission spectra reveals a typical linewidth one order of magnitude narrower than that of most samples reported in the literature. The narrowing of the PL line of L-SWNTs is due to a weaker effective exciton-phonon coupling subsequent to a weaker localization of the exciton. These results suggest that exciton localization in SWNTs not only arises from interfacial effects, but that the intrinsic crystalline quality of the SWNT plays an important role.
\end{abstract}

\maketitle
Photoluminescence (PL) emission in semiconducting carbon nanotubes arises from exciton recombination \cite{Wang2006, Dresselhaus2007, Voisin2012} and has been extensively studied in view of possible applications in opto-electronics, bio-imaging or photovoltaics \cite{Miura, Park2013, Welsher2009, Avouris2008}. Observation of photon antibunching in the near infrared \cite{Hogele08, Walden-Newman2012} suggests that SWNTs are also promising single-photon sources for the implementation of quantum information protocols. Interestingly, the PL emission energy (\textit{i.e.} the excitonic recombination energy) strongly depends on the tube diameter and can be easily tuned in the telecommunication bands at 0.83~eV ($1.5 \mu$m) by choosing SWNTs with a diameter of about 1-1.2 nm \cite{Lauret2003}. SWNTs could therefore make up a very versatile light source for quantum optics.
Several studies suggested that the optical properties of SWNTs at low temperature are best described in terms of localized excitons (zero-dimensional confinement), leading to a quantum dot like behavior \cite{Galland2008, Hofmann2013}. Nevertheless, the nature of the traps responsible for this exciton localization is not elucidated yet. In order to address the issue of exciton localization, we studied carbon nanotubes produced by high-temperature synthesis methods such as electric arc or laser ablation methods, which are known for their higher crystalline quality, with a lower density of defects \cite{Thess26071996, journet, Thostenson, Journet98, Journet12}. \\

In addition to the large Coulomb interaction responsible for the huge exciton binding energy in SWNTs, many other effects - both intrinsic (exciton - phonon coupling \cite{Htoon2005, Galland2008, Vialla}, exciton - exciton interaction\cite{Matsuda2008}, structural defects \cite{Georgi, Brozena2014},\dots) and extrinsic (residual doping \cite{Crochet2012}, interfacial inhomogeneities \cite{Ma2014}, fluctuating environment \cite{Ai2011, Walden-Newman2012},\dots) - are known to significantly alter the emission properties of SWNTs and act as dephasing sources for the exciton. The corresponding figure of merit is the full width at half maximum (FWHM) of the PL line that contains the optical signatures of exciton dephasing processes. Importantly, this line-width has to be measured at the single emitter scale to avoid inhomogeneous broadening \cite{Fwang}, but only the statistical analysis of the line-width distribution is relevant to benchmark the quality of a sample and thus of a growth/post-processing method.

In this work, we study the low-temperature (10~K) PL emission from surfactant dispersed laser ablation carbon nanotubes (L-SWNTs) \cite{Guo1995, Jost1999}. These nanotubes result from a high-temperature growth process and are well-known for having a higher crystalline quality. With a mean diameter of the order of 1.1~nm, they typically emit in the telecommunication band at about 1.5 $\mu$m \cite{Lauret2003}. Our sample is obtained by spinning the solution-processed SWNTs on the flat surface of a solid immersion lens coated with poly-lysine, following the previously reported procedure \cite{Hogele08}. All nanotube suspensions are obtained by dispersing raw powders in water with 2\% in weight of sodium cholate under vigorous sonication for one hour. The microphotoluminescence setup consists of a home-built confocal microscope, with a 0.4~NA microscope objective. The sample is excited by a laser diode emitting at 830 nm and the signal is then analyzed by a 50cm spectrometer equipped with a 600 l/mm grating blazed at 1.6 $\mu$m. A nitrogen cooled InGaAs array is used to record the spectra. By performing a statistical study on more than 40 SWNTs, we find a FWHM distribution centered around 500 $\mu$eV with values as low as 250 $\mu$eV (resolution limited). We stress that, despite the lack of control of the microscopic dielectric environment in our sample, the observed  FWHM distribution is centered at values that are smaller than most of the values presented in the
literature \cite{Htoon2004,Lefebvre2004,Matsuda2008}, or comparable to the values reported in studies focusing on the control of the environment-induced dephasing (air-bridging or polymer wrapped SWNTs) \cite{Ai2011,Sarpkaya2013}. As a reference for CVD-grown SWNTs, we compare typical spectra of L-SWNTs with those of similarly processed CoMoCat SWNTs. We show that the reduction of the effective exciton-phonon coupling due to a weaker exciton localization in L-SWNTs is the key factor for the narrowing of the PL lines. The emission profiles are fitted using the extended exciton-phonon coupling model \cite{Galland2008,Vialla}. We deduce that the exciton localization length $\sigma$\footnote{$\sigma$ is the extension of the exciton center-of-mass gaussian envelope wave function} is in the range 10-16~nm, which is up to $\sim$6~times larger than the value reported for CVD-grown SWNTs \cite{Galland2008,Vialla}. This suggests that the exciton localization in SWNTs is not enterly due to interfacial effects, but that the intrinsic crystalline quality of the SWNT matters.

\begin{figure}
\includegraphics[scale=0.30]{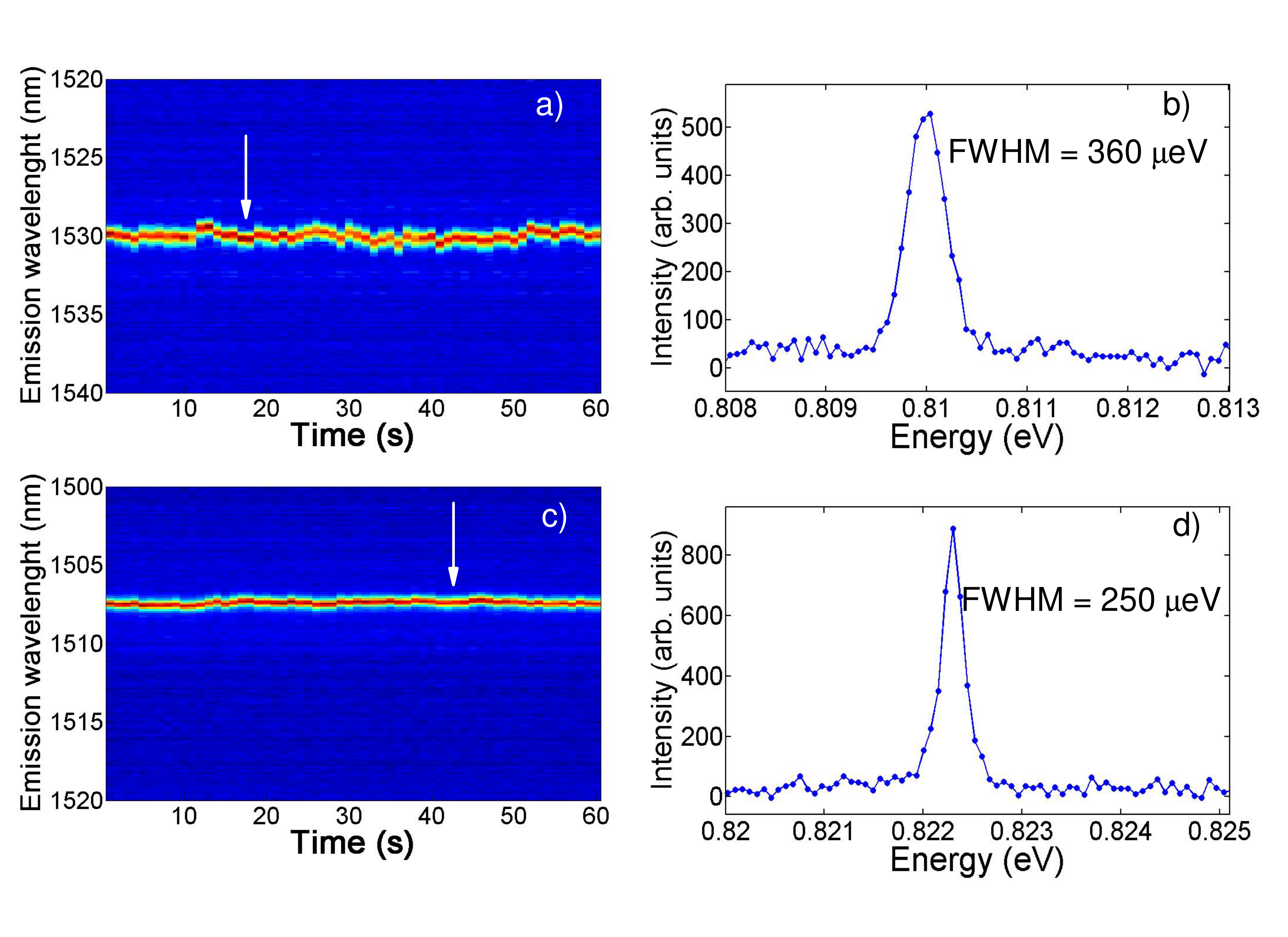}
\caption{a) and c) Temporal traces of the PL emission of two different SWNTs consisting of 60 consecutive spectra (1 second integration time for each spectrum). b) and d) Example of a spectrum extracted from the trace in a) and c) (indicated by the white arrows). The power density used here is 60 $\mathrm{kW/cm^2}$.}
\label{fig1}
\end{figure}

Figure \ref{fig1} (a,c) shows a sequence of 60 spectra (acquisition time of 1 second for each spectrum) for two typical individual SWNTs found in our sample, with emission wavelengths of 1530 nm and 1507 nm respectively. These temporal traces show the usual features of PL from individual surfactant dispersed SWNT, that are spectral diffusion (SD) and blinking. In figure 1 (b,d) two spectra extracted from the sequence are represented. They show a FWHM of 360 and 250 $\mu$eV respectively.
In presence of SD, the measured FWHM can be considered as an upper limit to the intrinsic FWHM. In the following, we study this upper limit by setting the acquisition time to 1 second. This value ensures a good signal to noise ratio for a great number of SWNTs in our sample.
Moreover, no significant change in FWHM is observed in the  pump power range used in this work, and the PL intensity shows a linear dependence on the incident power (See Supplemental Material for further details \footnote{See Supplemental Material for further details at}). This suggests that exciton-exciton interactions \cite{Matsuda2008} give a negligible contribution to the measured FWHMs.\\

\begin{figure}
\includegraphics[scale=0.25]{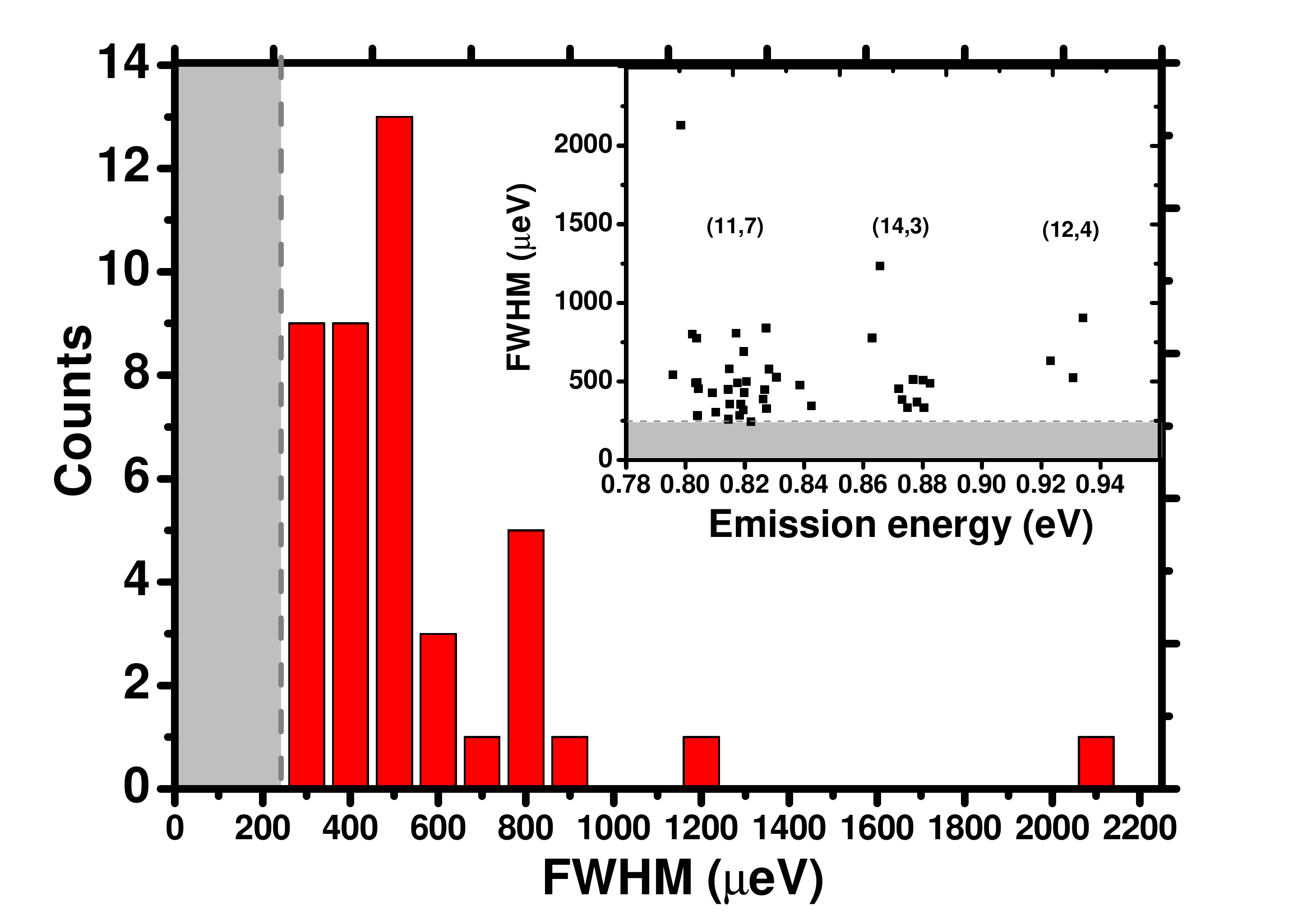}
\caption{FWHM distribution measured for 43 individual SWNTs. The data reported on the histogram are obtained by picking the minimum FWHM in the temporal traces such as those showed in figure \ref{fig1}. The most probable values are around 500 $\mu$eV. The low-values tail is cut off by the optical resolution of the setup at 250 $\mu$eV (gray dashed line). Inset : dependence of FWHM on emission energy.}
\label{fig2}
\end{figure}

In order to check if L-SWNTs show a statistically narrower FWHM than previously reported samples, we measured the FWHM distribution. We analyzed 43 SWNTs, resulting in the histogram shown in figure \ref{fig2}. The FWHM distribution is centered around 500~$\mu$eV, with a lower (resolution limited) FWHM of 250~$\mu$eV. The asymmetry of this distribution, due to instrumental constraints, strongly suggests that even narrower PL lines could be observed in this sample. This FWHM distribution shows that L-SWNTs have a narrower PL emission than most solution processed SWNTs  which show typical FWHMs of several meV \cite{Sarpkaya2013, Hofmann2013, Ma2014,Htoon2004}. Strikingly, these FWHM values seem comparable to and even narrower (at least in term of distribution) than those observed for most air-bridging CVD SWNTs \cite{Sarpkaya2013} and for most neutral polymer-embedded SWNTs \cite{Ai2011,Walden-Newman2012, Alexander-Webber2014}, \textit{i.e.} in samples in which the fluctuations of the microscopic dielectric environment are reduced.

 The FWHM distribution of figure \ref{fig2} results from a complex interplay of different broadening phenomena. In the case of carbon nanotubes, FWHM measurements essentially encompass two different mechanisms, namely the so-called Ohmic broadening and the pure dephasing broadening. The former corresponds to the non-perturbative coupling of a localized exciton to a 1D phonon bath, that is known to give rise to a broad and asymmetric PL lineshape \cite{Galland2008, Vialla}. The latter gives an additional contribution to the global FWHM and becomes the dominant dephasing mechanism when the Ohmic broadening is suppressed. The intermediate case known as extended exciton-phonon coupling model accounts for all the distinct PL profiles observed at low temperature, consisting of a sharp zero-phonon line (ZPL) with phonon side-bands of variable magnitudes \cite{Vialla}. In the following, we show that the reduction of the effective exciton-phonon coupling subsequent to a larger exciton delocalization plays a key role in the interpretation of our experimental findings. \\

\begin{figure}
\includegraphics[scale=0.3]{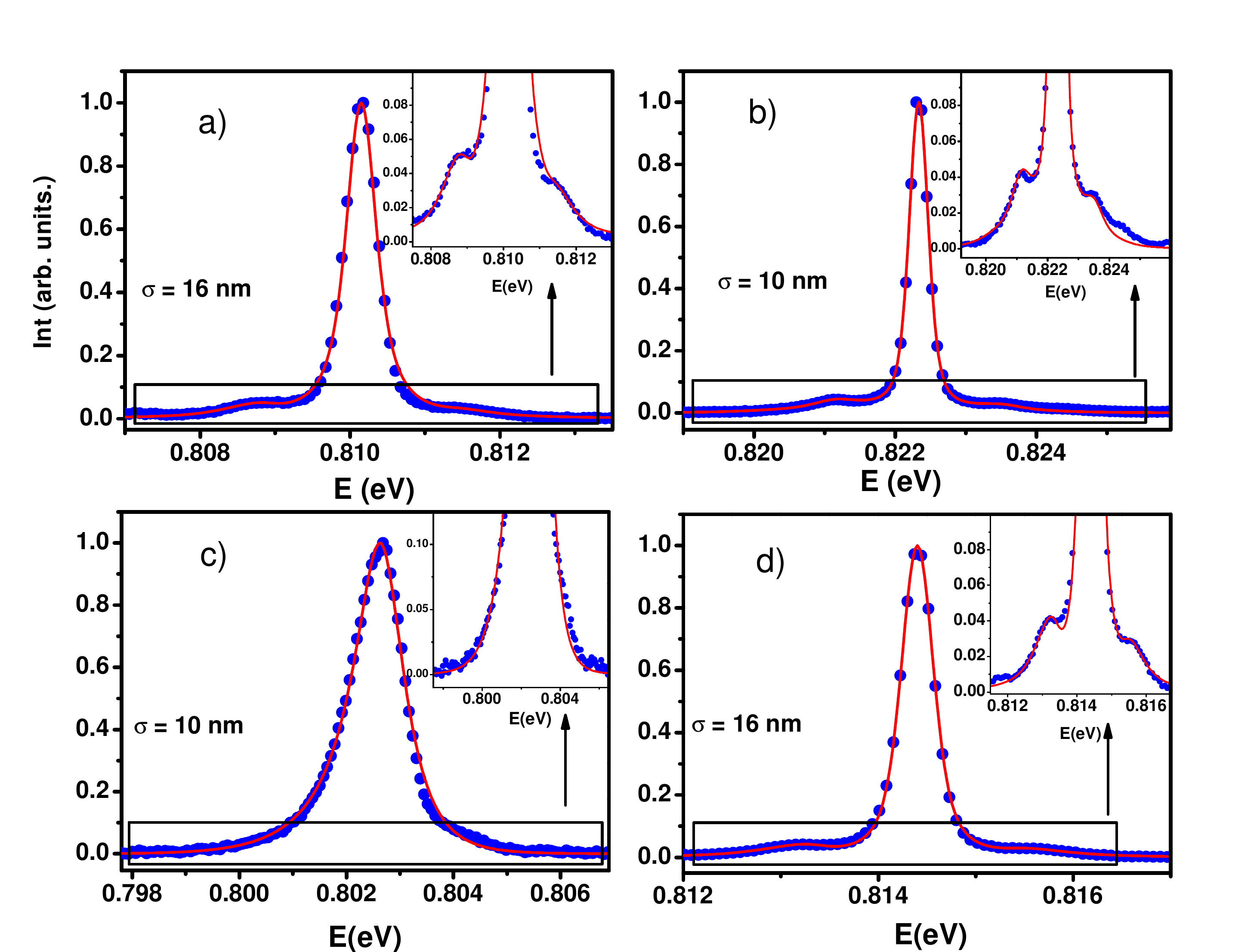}
\caption{PL spectra of L-SWNTs : a), b) are the tubes of figure \ref{fig1}. a), b) and d) were acquired with an integration time of 60s and show small phonon wings on either sides of the ZPL. In c) a broader profile with an asymmetric shape (Integration time 20 s). In red, fits to the experimental data using the non perturbative exciton phonon coupling model (see main the text and SI). Insets: magnified view on the phonon sidebands.}
\label{fig3}
\end{figure}

\begin{figure}
\includegraphics[scale=0.25]{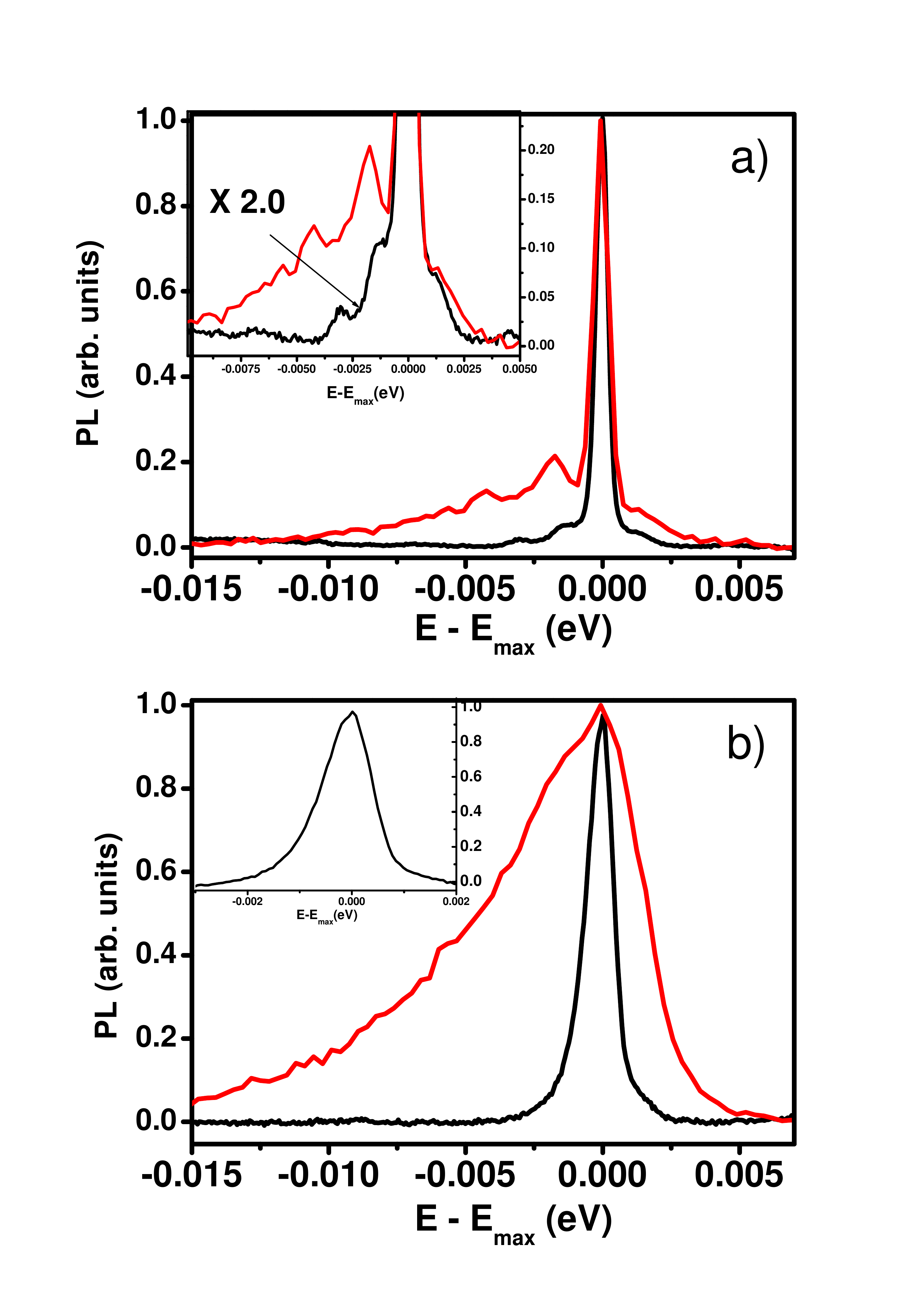}
\caption{In black two typical PL spectra of L-SWNT. In red PL spectra of CoMoCAT SWNT with the same type of line profile. a) Exemple of sharp ZPL emerging from the phonon wings. In the inset : zoom on the phonon wings. b) Example of triangular asymmetric shape. In the inset : zoom on the L-SWNT PL spectra to highlight the asymmetric shape. The energy axes are shifted in order to superimpose the PL peaks.}
\label{fig4}
\end{figure}

Figure \ref{fig3} displays high signal to noise ratio spectra recorded in L-SWNTs. Note that despite the 60-fold increase of the integration time used to acquire these high-resolution spectra, the extra broadening is on the order of 100-200~$\mu$eV at most, which remains small as compared to the extension of the typical spectral features (several meV), showing that spectral diffusion is not the dominant broadening mechanism. The observation of all the representative PL profiles predicted by the extended exciton-phonon coupling model shows that the diversity of line-profiles reported in the literature \cite{Vialla} is not specific to the case of CVD nanotubes but is also encountered in L-SWNTs. To highlight this point, we compare in Figure \ref{fig4} some examples of PL spectra of L-SWNTs with those of CoMoCat nanotubes showing an identical line-shape. The profile similarity is striking except that the line-width is consistently lower for L-SWNTs.

This means that the nature of the broadening mechanisms is similar in all types of nanotubes but that their magnitude is reduced in L-SWNTs. In this respect, it is worth noting that the typical SD magnitude (as measured by the standard deviation of the peak energy in a series of 60 spectra acquired with a 1~s integration time) is of the same order for CoMoCat and L-SWNTs (e.g. $\sim$~100$\mu$eV for the tubes of figure \ref{fig4}-a), ruling out that SD is the source of the difference between the two types of sample.

Spectra a), b)and d) of figure \ref{fig3} show the presence of a pedestal formed by small asymmetric phonon wings on either sides of the ZPL. In c) the ZPL is merged in the phonon wings, forming an overall broader and asymmetric PL line-shape. The red lines in figure \ref{fig3} are the fits obtained with the extended exciton-phonon coupling model \cite{Vialla, Galland2008}(See Supplemental Material for the details).
 Here, we briefly recall that for a localized exciton coupled to a truly 1D phonon bath, the exciton-phonon coupling diverges for the low-energy phonon modes leading to a typical broad and asymmetric lineshape. In contrast, when the local density of states of acoustic phonons at the exciton location is reduced (\textit{e.g.} by a local mechanical coupling to the environment), a sharp ZPL is restored and emerges from the phonon wings.
Most importantly, in both cases the extension of the phonon sidebands directly mirrors the size of the (Gaussian) exciton envelope $\sigma$ used to describe the  localization of the exciton center of mass. In fact, a finite value of $\sigma$ corresponds to a finite extension of the exciton wave-function in the $k$-space leading to a cut-off energy of the exciton-phonon coupling at $\Delta E_{\sigma} =  2 \hbar v/\sigma$ \footnote{Here $v=19.9$  km/s is the sound velocity for the effective phonon mode obtained by merging low energies stretching and twisting phonon branches  \cite{Suzuura2002, Nguyen2011}}, due to the momentum conservation in the exciton-phonon interaction. Therefore, the spectral width of the phonon sidebands is a probe of the exciton localization. The values of $\sigma$ extracted from the fits (red curves in figure \ref{fig3}) are reported on the graphs. Interestingly, all the tubes studied here show values of $\sigma$ in the range 10 - 16 nm. These values are larger than the values found for CoMoCat tubes with the same fitting model ($\sim$3nm) \cite{Vialla,Galland2008}. Large values of $\sigma$ are hallmarks of weaker exciton localization in L-SWNTs, which we interpret as a consequence of the higher crystalline quality of this material. This shows that several degrees of exciton localization are achievable in SWNTs. Therefore, identification of high crystalline quality nanotubes is a key step towards the control of exciton localization, going continuously from 0D to 1D confinement.

Previous studies reported a diameter dependence of the magnitude of the electron-phonon coupling \cite{Yoshikawa2009, Inoue, Suzuura2002}. In fact, the increase of the tube linear mass density, which is proportional to the tube diameter, also gives rise to a reduction of the exciton-phonon coupling \cite{Nguyen2011}.  In order to see if such diameter related effects are visible in our data, we checked for a possible correlation between the FWHM and the energy of the PL emission. In the inset of figure \ref{fig2}, the FWHM of the PL spectra are plotted as a function of the emission energy. Three groups of points are visible and are tentatively assigned to (11,7), (14,3), (12,4) chiralities \cite{Lauret}.
Due to the scattering of the experimental points and due to the poor statistics for the smaller diameters, it is hard to identify any clear trend. Moreover, we note that previous studies with CVD nanotubes in the same range of diameters led to much larger linewidths ruling out the diameter dependence of the electron-phonon coupling as a main source of narrowing of the PL spectra for L-SWNTs \cite{Lefebvre2004}.

Finally, we discuss the possible origin of the residual dephasing in SWNTs with prominent ZPL (non Ohmic exciton-phonon coupling). This broadening can be due to residual SD on a time-scale smaller than the integration time. Qualitatively, we do observe a correlation between the FWHM and the SD measured at the 1 s time-scale (See Supplemental Material). With a contribution of the order of 100 $\mu$eV as compared to an average FWHM of 500~$\mu$eV, SD plays a non-negligible role on the ZPL line-width. This is exemplified in the spectra of Figure 4-a, where the L-SWNT and the CoMocat spectra show comparable ZPL widths, which is consistent with the comparable magnitude of SD measured in each tube.

In conclusion, we studied the low-temperature PL emission from laser ablation solution-processed SWNTs emitting in the telecommunication band around 1.5~$\mu$m. A statistical analysis shows a distribution of line-widths centered around 500~$\mu$eV, which is significantly narrower than the previously reported values for solution-processed nanotubes. Moreover, L-SWNTs show a diversity of PL profiles which is very similar to that of CVD nanotubes, showing that their emission spectra is also driven by the exciton-phonon coupling. However, the consistently weaker and narrower phonon wings point to an up to 6-fold larger localization length of the excitons in L-SWNTs, as deduced from a fitting of the line profiles to the extended exciton-phonon coupling model. This finding shows that the exciton localization at low-temperature is not governed by the sole interfacial coupling to the environment but also to a large extend by the intrinsic crystalline quality (\textit{i.e.} the defect density) of the nanotube.
Finally, it is worth noting that this study reveals that several degrees of exciton localization are achievable in SWNTs at low temperature, ranging from nearly 0D quantum dot like excitons in CVD-nanotubes to a pronounced 1D flavor in L-SWNTs with localization lengths exceeding 10 times the exciton Bohr radius. Identification of high crystalline quality SWNTs is therefore a key issue since it could provide a model system for the study of confinement effects in one-dimensional excitons going continuously from near one-dimensional excitons to truly zero-dimensional fully confined excitons. Finally, it may open the way to the control of the exciton localization, an important step towards the realization of deterministic single photon source at telecommunication wavelengths based on carbon nanotubes. \\

This work is supported by a public grant overseen by the French National Research Agency (ANR) as a part of the "Investissements d'Avenir" program (reference: ANR-10-LABX-0035, Labex NanoSaclay), by the GDR-I GNT, the ANR grant TRANCHANT.  C.V. is a member of Institut Universitaire de France. Laser ablation SWNTs were provided by O. Jost (now at Fraunhofer IWS).

\bibliography{./Biblio_aps}

\end{document}